\documentclass[twocolumn,preprintnumbers,amsmath,amssymb,superscriptaddress, altaffilletter, subeqn,prl, maxbibnames=99]{revtex4-2}
\usepackage{physics}
\usepackage{dcolumn}
\usepackage{bm}					% Bold Greek letters in math mode.
\usepackage{graphicx}
\usepackage{amsmath}
\usepackage[dvipsnames]{xcolor}
\usepackage{comment}
\begin{document}

%\title{Ab initio exciton dynamics in TMD heterostructures: from  absorption to emission}
\title{Ultrafast Exciton Decomposition in Transition Metal Dichalcogenide Heterostructures}

\author{Tomer Amit}

\affiliation{Department of Molecular Chemistry and Materials Science, Weizmann Institute of Science, Rehovot 7610001, Israel}

\author{Sivan Refaely-Abramson}
 
\affiliation{Department of Molecular Chemistry and Materials Science, Weizmann Institute of Science, Rehovot 7610001, Israel}

\email[Corresponding author:]{sivan.refaely-abramson@weizmann.ac.il}

\begin{abstract}
Heterostructures of layered transition metal dichalcogenides (TMDs) host long-lived, tunable excitons, making them intriguing candidates for material-based quantum information applications. Light absorption in these systems induces a plethora of optically excited states that hybridize both interlayer and intralayer characteristics, providing a distinctive starting point for their relaxation processes, in which the interplay between generated electron-hole pairs and their scattering with phonons play a key role. We present a first-principles theoretical approach to compute phonon-induced exciton decomposition due to rapid occupation of electron-hole pairs with finite momentum and opposite spin. Using the MoSe$_2$/WSe$_2$ heterostructure as a case study, we observe a reduction in the optical activity of bright states upon phonon scattering already in the first few femtoseconds proceeding the photoexcitation, driving exciton interlayer delocalization and subsequent variations in the exciton spin. Our results reveal an unexpected and previously unexplored starting point for exciton relaxation dynamics, suggesting increased availability for coherent interactions and non-radiative processes through ultrafast changes in exciton momentum, spatial, and spin properties upon light excitation.
\end{abstract} 
 
\maketitle

Exciton relaxation processes underlying excited-state dynamics in heterostructures of transition metal dichalcogenides (TMDs) are a topic of broad interest~\cite{rivera2018interlayer, miller2017long,  jin2018ultrafast}. The combination of optically-active intralayer excitons, Coulomb-bound electron-hole pairs residing mainly within the individual layers, and low-lying interlayer excitons spread between the layers~\cite{rivera2015observation, jauregui2019electrical, Merkl2019, schmitt2022formation, karni2022structure}, induces relaxation mechanisms which are challenging to capture within simple frameworks. These involve a complex interplay between excitations with varying levels of layer and valley decomposition~\cite{forg2019cavity,kundu2022exciton}, determining spatial and spin properties that are directly coupled to system structure through the atomistic details of the participating layers and their relative alignment~\cite{barre2022optical, naik2022intralayer, sood2023bidirectional}. Gaining a comprehensive understanding of the involved excitonic processes and their structural dependencies is intriguing. In particular, considering variations in the excitonic nature along its time evolution subsequent to photoexcitation can guide design principles for effective generation of long-lived and ideally coherent exciton phases in TMD heterostructures~\cite{fogler2014high, wang2019evidence, sigl2020signatures, bai2022evidence, slobodkin2020quantum, troue2023extended, katzer2023exciton, mak2018opportunities, kennes2021moire}.

A common approach to accurately describe structure-specific exciton properties is by constructing an excitonic basis set through the solution of a many-body Bethe Salpeter equation within the so-called GW-BSE approach~\cite{hybertsen1986electron, rohlfing2000electron, *rohlfing1998electron}. Within this first-principles framework, Coulomb and exchange interactions between multiple electron and hole wavefunctions are computed by numerically accounting for a large number of participating bands, with varying crystal momentum and spin properties, computed explicitly. This allows for an accurate consideration of dielectric effects, spin-orbit coupling, and exciton dispersion~\cite{louie2005quasiparticle, louie2021discovering, qiu2015nonanalyticity, Cudazzo2015}. Further methodological advances enable computations of GW-based non-equilibrium effects on particle propagation by accounting for temporal evolution of the dielectric function and its consequences in time-resolved photexcitation processes~\cite{attaccalite2011real, perfetto2015nonequilibrium, sangalli2018ab, chan2021giant, sangalli2021excitons, perfetto2023real}. Still, taking this wealth of participating particles into account in the relaxation processes is an extremely challenging task. Recent computational advances and theory developments allow predictive calculations of exciton-exciton scattering through phonons~\cite{antonius2022theory, chen2022first, chan2023exciton, cohen2023phonon}. However, scattering with phonons can alter the internal exciton nature already at the single-exciton level~\cite{Paleari.2022}. Such effects can induce rapid changes in the exciton band composition and are thus expected to play a crucial role in the relaxation processes, in particular by modifying the optical selection rules that determine the crossplay between exciton radiative recombination and non-radiative exciton-exciton scattering.

In this letter, we present an \textit{ab initio} computational scheme to compute ultrafast exciton decomposition following light excitation. We calculate the exciton time evolution upon coupling to phonons within a first-principles-based density matrix formalism, in which the electron-hole pairs composing the excitons scatter simultaneously, leading to time-resolved changes in the pair population.  Demonstrated for the case of a MoSe$_2$/WSe$_2$ heterostructure, we show that such phonon-induced modifications in the electron-hole pair composition occur already within a few femtoseconds after excitation. These allow the occupation of momentum- and spin-indirect electron-hole pairs almost immediately, leading to instantaneous changes in the exciton oscillator strengths due to the induced layer delocalization and population of opposite-spin transitions. Our results demonstrate the delicate effects of ultrafast scattering events on the exciton properties, suggesting a new understanding of the starting point for exciton relaxation dynamics and shedding light on coherent coupling mechanisms between optically-active intralayer excitons and long-lived interlayer excitons in TMD heterostructures.

 \begin{figure*}
\includegraphics[width=1.0\linewidth]{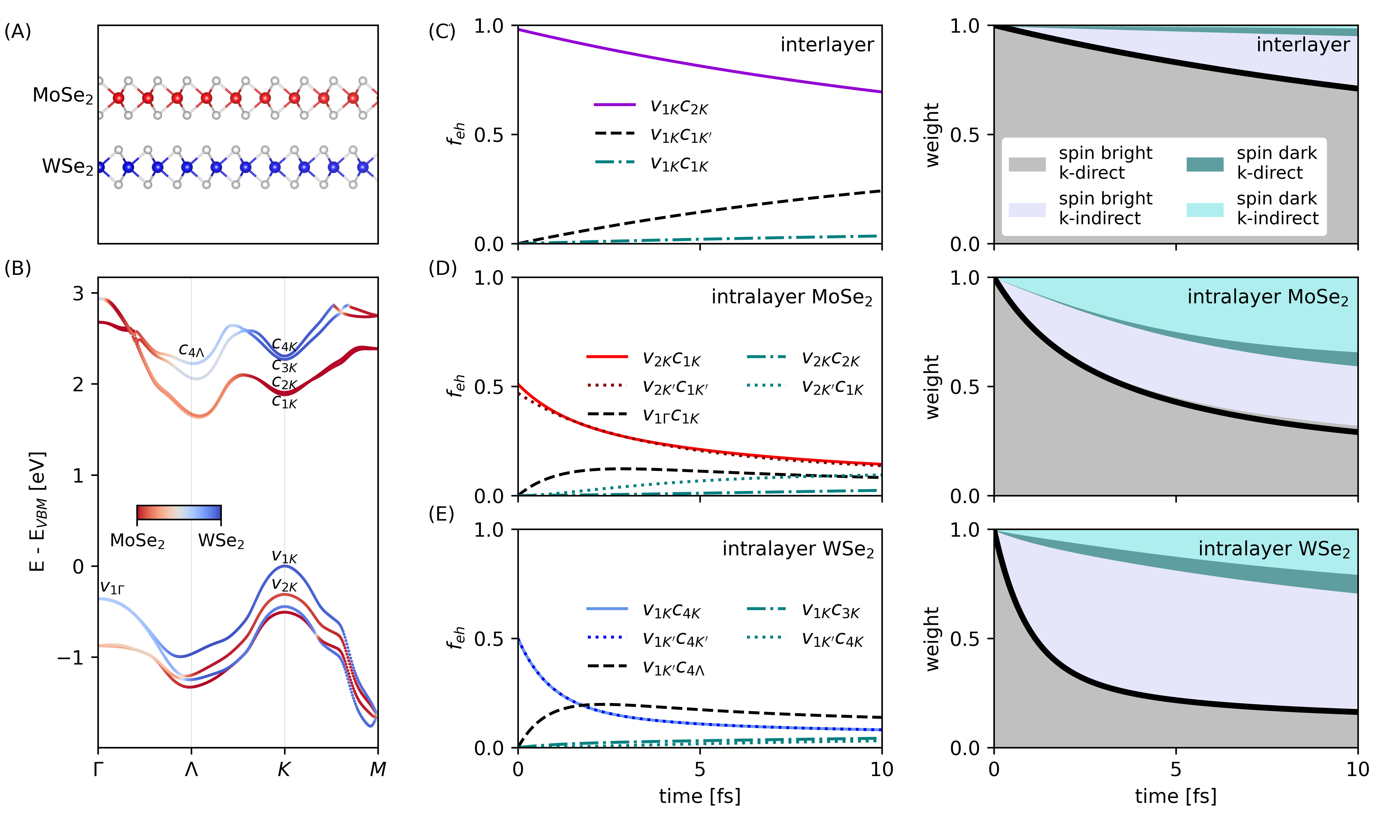}    
    \caption{Phonon-driven electron-hole pair decomposition of intelayer and intralayer exciton states in the MoSe$_2$/WSe$_2$ heterostructure. (A) Schematic representation of the examined TMD heterostructure of commensurate MoSe$_2$/WSe$_2$ in H$_h^h$ stacking. (B) Electronic quasiparticle bandstructure computed within the GW approximation. Band colors represent the relative MoSe$_2$ (red) and WSe$_2$ (blue) layer contribution within the bands. Conduction (c) and valence (v) regions are shown with the notation of $v_1$, $c_1$ for the highest valence and lowest conduction bands, respectively, and so forth. (C-E) Time-resolved evolution of three representative exciton states: interlayer, intralayer MoSe$_2$, and intralayer WSe$_2$. The initial occupation at time $t=0$ represents the GW-BSE solution. Evolution of the relative electron-hole pair population in these excitons due to phonon scattering is shown in the left panels, for selected contributing pairs $v_{im}c_{jn}$ of holes and electrons at bands $i,j$ and k-points $m,n$, respectively. The right panels show the weighted pair occupation, summed over all participating transitions. The initially-excited pairs are optically-bright, with time-resolved population of momentum-indirect pairs, spin-dark pairs, and combination of the two. Black line represents the relative oscillator strength upon pair decomposition. \label{fig:dynfig1}}
\end{figure*}

We study the time evolution of optically-excited low-energy interlayer and intralayer excitons in the MoSe$_2$/WSe$_2$ heterostructure with H$_h^h$ layer alignment, schematically shown in Fig.~\ref{fig:dynfig1}(A). Our starting point includes an electronic-structure evaluation of the electronic and excitonic states in this system, computed within the GW~\cite{hybertsen1986electron} and GW-BSE~\cite{rohlfing2000electron, *rohlfing1998electron} approximation built on top of density functional theory (DFT)~\cite{kohn1965self}. In this approach, exchange and screened Coulomb interactions between electrons and holes are computed through a full evaluation of the dielectric function and accounting for explicit wavefunction coupling of both the spatial and spinor parts. At heterostructures, this procedure results with multiple optically-excited states, with absorption structure strongly depending on the underlying heterostructure composition and alignment~\cite{barre2022optical, hernangomez2023reduced, kundu2022exciton}. Phonon modes are computed within density functional perturbation theory (DFPT)~\cite{giustino2017electron}, allowing a first-principles assessment of the participating electron-phonon and hole-phonon coupling (see SI for full computational details).

Figure~\ref{fig:dynfig1}(B) shows the calculated GW quasiparticle band structure, with band colors representing the relative wavefunction contribution from the MoSe$_2$ (red) and WSe$_2$ (blue) layers. We first focus on the low-lying interlayer and intralayer exciton states, photoexcited around the K/K' valleys. The electron-hole transitions dominating the interlayer excitation is from the valence band ($v_1$), primarily localized on the WSe$_2$ layer, to the conduction band (a spin-split of $c_{1,2}$), mainly localized on the MoSe$_2$ layer. Intralayer MoSe$_2$ transitions are primarily from the second valence band ($v_2$) to the conduction region, and intralayer WSe$_2$ transitions from $v_1$ to the higher-energy spin-split conduction bands $c_{3,4}$. As a first step, we analyze the change in the electron-hole pairs composing these states and show that intralayer excitons inherent an interlayer nature within a few femtoseconds upon coupling to phonons. 

We compute the time evolution of the electron-hole pairs composing the excitons within a Lindblad-type density-matrix representation~\cite{rosati2014derivation, rosati2015electron, xu2020spin}.   
We define an initial state of an optically-bright exciton, as computed from GW-BSE. This sets an initial density matrix that can generally be composed of the various participating electron-hole pairs within the exciton basis set, \begin{equation}
\begin{split}
 &\hat{\rho}\left(t=0\right)=\left|S\right>\left<S\right|\\
 &\left|S\right>=\sum_{ehkQ}A_{ehkQ}^{S}\ket{h,k}\ket{e,k+Q}.
 \end{split}  
\end{equation}
Here, $\left|S\right>$ is an eigenstate solving the GW-BSE, with $|A^S_{ehkQ}|^2$ the probability amplitudes for each pair transition between holes ($h$) with crystal momentum $k$ and electrons ($e$) with crystal momentum $k+Q$. The time propagation is computed by allowing each electron-hole pair to interact independently with phonons. The underlying assumption for such interaction is that the optical exciton basis, generally composed of various electron-hole transitions energetically resonating together, can alter in the relaxation process. We thus explore the change in pair composition while allowing the hole and electron states to scatter into other bands through phonons. Room temperature is assumed throughout. In practice, this sets time-dependent changes in the occupation of the $|A^S_{ehkQ}|^2$ coefficients. While all initially occupied electron-hole pairs on the diagonal of $\hat{\rho}\left(t=0\right)$ are assumed to have momentum $Q=0$ within the optical excitation, upon phonon scattering, finite $Q$ states become occupied, inducing layer mixing and allowing subsequent population of opposite-spin bands. This multi-step propagation leads to rapid occupation of dark electron-hole transitions, effectively reducing the exciton oscillator strength, as we demonstrate below. 
  
Figure~\ref{fig:dynfig1}(C), left panel shows the phonon-induced changes in the occupation $f_{eh}$ of the main electron-hole ($eh$) pairs composing the interlayer exciton in the first 10 femtoseconds proceeding photoexcitation. The initial occupation of the chosen state results from the $v_1$ to $c_2$ electron-hole transition at the $K$ point, namely $v_{1K}c_{2K}$. Notably, the computed GW-BSE oscillator strength of this transition is only two orders of magnitude smaller than the optically-active intralayer excitons. This supports recent observations of optically-allowed interlayer excitons in this systems~\cite{barre2022optical, forg2019cavity, wietek2023non}. The pair occupation is modified already in the first few femtoseconds due to a rapid intervalley transition into the $v_{1K}c_{1K'}$ pair (black dashed line). This transition is a result of electron-phonon interactions between the two spin-like conduction states, $c_{2K}$ and $c_{1K'}$, at the opposite valleys $K,K'$. We note the difference between this phonon-induced ultrafast transition of exciton partial occupation and previously-explored exchange-driven valley dephasing, expected to occur at longer timescales~\cite{glazov2014exciton}. Following this transition, population of the pair $v_{1K}c_{1K}$ (teal dashed-dotted line) further occurs. This corresponds to an intravalley transition in which electrons scatter between the spin-split conduction bands $c_{2K}$ and $c_{1K}$. This transition is somewhat unexpected, but easy to understand when considering the broken spinor symmetry in this structure, as we demonstrated in previous work~\cite{barre2022optical}. The right panel summarizes the overall change in the electron-hole pair composition when summing over all participating pairs in the density matrix evolution. The initial pair occupation, momentum direct and spin bright by definition, quickly induces fractional population of momentum indirect and spin dark contributions. 

Figures~\ref{fig:dynfig1}(D,E) show a similar analysis for two bright intralayer excitons, both dominating the corresponding peak regions in the computed absorption (see SI). In these states, the initial excitation is split between the two valleys (we note that this representation is invariant and can in general modify upon unitary transformation of degenerate solutions of the GW-BSE). Importantly, unlike the interlayer case, the immediate phonon relaxation of these states now includes transitions outside of the $K/K'$ valleys. For example, for intralayer MoSe$_2$, fast occupation of electron-hole pairs with the hole residing on the $\Gamma$ point and the electron on the $K/K'$ point are populated almost instantaneously (dashed black line). As an immediate, rather surprising result, this population follows with rapid occupation of pairs with dark spin (dashed teal lines). The corresponding substantial changes in the time-resolved weighted occupation of momentum-indirect and spin-dark states is shown in the right panel. We note that the fast occupation of optically-dark states is a direct outcome of the participation of hole states at $\Gamma$ in the density matrix scattering dynamics, a point in which the electron-hole spin number is ill-defined. In the case of an initial intralayer WSe$_2$ exciton, rapid occupation of pairs with electrons residing at the $\Lambda$ point dictates a large change in the momentum directness of the excited state, with population of indirect, optically-dark transitions exceeding direct ones already within the first few femtoseconds. Again, these transitions further induce population of spin-flip states. These are slightly reduced compared to intralayer MoSe$_2$ upon the relatively-conserved spin at the $\Lambda$ point. We emphasize that these spin properties are highly sensitive to the band character, which can be structurally modified, for example, through interlayer twisting~\cite{kundu2022exciton, naik2022intralayer}.

\begin{figure*}
 \includegraphics[width=1.0\linewidth]{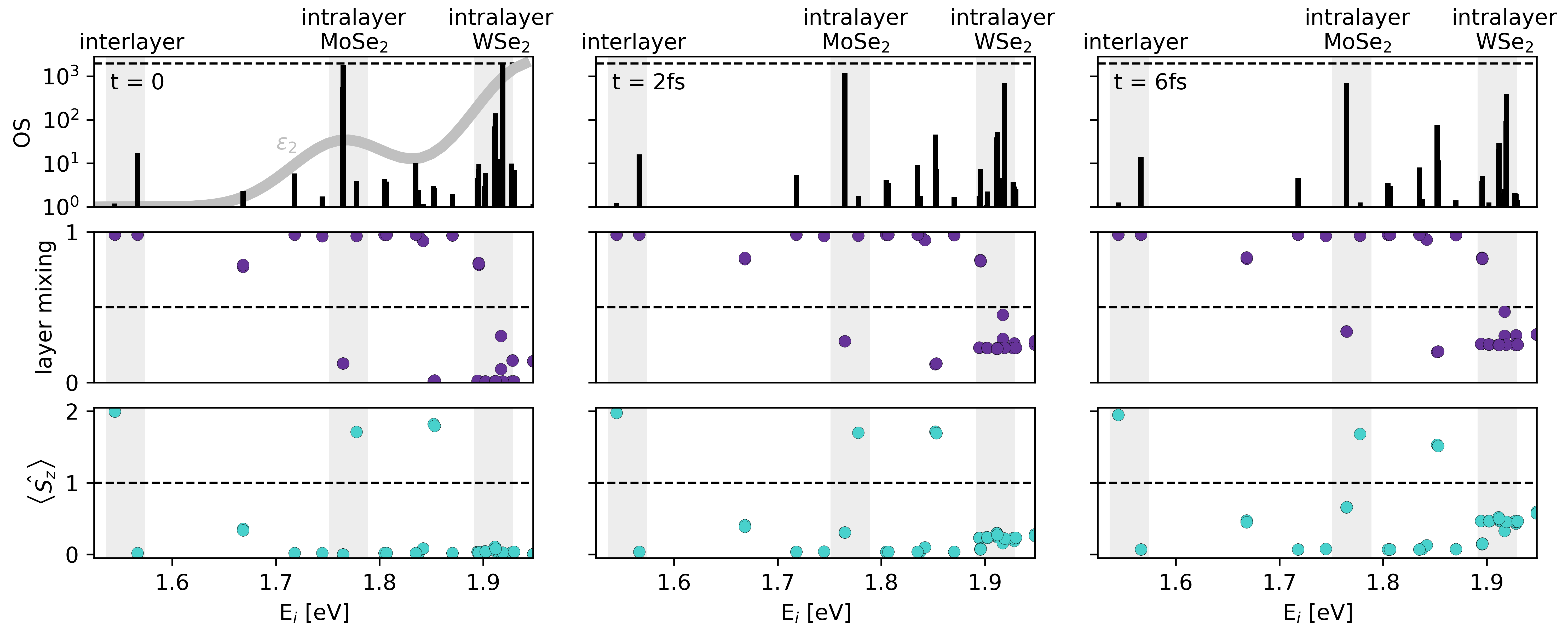}    
    \caption{Time-resolved evolution of the bright exciton states in the examined heterostructure. Top panels shows the oscillator strength for each exciton in a logarithmic scale, computed through the oscillator strengths of single electron-hole pairs and their varying relative contributions. The computed GW-BSE absorption spectrum at $t=0$ is shown in the top left panel (grey line). Middle panels show the change in layer mixing for each state, with $0$ representing fully intralayer and $1$ representing fully interlayer excitons. Bottom panels show the corresponding out-of-plane spin expectation value for each state, where $0$ stands for spin-like and $2$ for spin-unlike exciton transitions. Reduction in the oscillator strengths of the intralayer states occurs due to an increase in layer mixing, with associated population of spin-dark states, considerably modifying the starting point of the exciton relaxation dynamics.}\label{fig:dynfig2} 
\end{figure*}

While the above-chosen interlayer and intralayer states illustrate the origins of exciton decomposition in our approach, the computed GW-BSE absorption spectrum includes a much larger number of bright excitations ranging between the interlayer and intralayer excitons, with varying initial compositions of electron-hole pairs (see SI). To capture the collective effect of the above-discussed time-resolved modifications, Fig.~\ref{fig:dynfig2} shows the phonon-induced changes along this energy range. We specifically examine the time evolution of the exciton oscillator strength (OS), calculated through the single-pair oscillator strengths and their time-dependent populations; the level of layer mixing, where $0$ stands for pure intralayer and $1$ for pure interlayer states; and the out-of-plane spin expectation value $\braket{S_z}$, where $0$ describes spin-like and $2$ spin-unlike transitions. We present only initially-bright states with computed OS$>1\ e^2a_0^2$. The energy regions with the low-lying interlayer and intralayer states explored above are marked with grey shaded areas.

The immediate effect of the phonon-induced pair occupation is directly observed through the reduction in the OS. This change is particularly visible at both intralayer regions, where the OS of the brightest transitions is reduced by an order of magnitude. In addition, we note the immediate OS reduction of states surrounding the main intralayer excitons, effectively suggesting a decreased broadening of these peaks. We note that this is still a single-exciton property; further broadening effects are expected upon exciton-exciton interactions, not included here. The associated changes in layer mixing and spin multiplicity demonstrate that these reductions in the OS are assigned to occupation of electron-hole pairs that enhance the interlayer nature of these excitons, involving states outside of the $K,K'$ valleys. 

The calculated pair decomposition and rapid layer mixing suggest that the exciton basis set computed for photoexcited states in absorption is not necessarily kept within the relaxation. These observations supply a complementary view of recent findings by Paleari et al.~\cite{Paleari.2022}, in which changes in the exciton state were computed upon simultaneous coupling of electrons, holes, and phonons within a self-consistent extended-BSE kernel, composing the so-called \textit{elemental} excitons, as opposed to \textit{optical} ones. Using the first-principles density matrix approach we presented here, and under the assumption of weak particle-phonon coupling, our findings offers an alternative view into the origins of these changes. We note that additional energy relaxation of the exciton state is expected to occur due to time-resolved modifications in the dielectric screening and corresponding changes in the electron-hole coupling. Such effects were recently studied in TMD monolayers for the non-excitonic case~\cite{perfetto2023real}, suggesting that these only become significant at longer timescales than the ones examined here.   

The presented approach for computing the exciton-phonon interaction dynamics is closely related to the common assumption that exciton-phonon scattering is primarily dominated by the static Fan-Migdal interaction terms, in which electron-phonon and hole-phonon coupling are weighted within the coupled exciton-phonon terms through GW-BSE state components~\cite{antonius2022theory}. While recent studies employed this approach to compute exciton-phonon scattering rates~\cite{chen2022first, chan2023exciton, cohen2023phonon}, our approach here is different: we keep the interactions within a single-exciton picture, but allow a change in the electron-hole pairs composing the excitons in the BSE basis set.  Within this representation, this exciton basis set of the optical excitations is effectively broken, suggesting a loss of exciton state orthogonality which can enhance coherent transitions and strong Coulomb coupling between the newly generated states. In particular, the layer mixing occuring directly after excitation of intralayer states implies on effective coupling to long-lived interlayer excitons. We leave the numerical exploration of these exciton-exciton interactions and an evaluation of the anticipated exciton coherences to future work. 

In conclusion, we present a first-principles approach to compute exciton ultrafast dynamics through phonon-induced electron and hole pair occupation following an optical excitation. 
Our analysis, based on an \textit{ab initio} density matrix formalism, points to a rapid reduction in the optical activity of intralayer excitons due to layer mixing. Furthermore, it suggests a new perspective of the starting point for the relaxation between optically-excited intralayer states and long-lived interlayer states, with spatial overlap between these excitons induced almost immediately after excitation, followed by occupation of spin-dark components.
Our findings supply a structure-specific and state-dependent measure of the exciton decomposition proceeding photoexcitation, facilitating a predictive understanding of exciton relaxation due to phonon scattering from a bandstructure perspective.\\

\noindent{Acknowledgments:}
We thank Ouri Karni and Uri Peskin for insightful discussions. S.R.A. is an incumbent of the Leah Omenn Career Development Chair and acknowledges a Peter and Patricia Gruber Award and an Alon Fellowship. T.A. is supported by the David Lopatie Fellows Program.
The project has received further funding from the European Research Council (ERC), Grant agreement No.101041159, and an Israel Science Foundation Grant No.1208/19. Computational resources were provided by the ChemFarm local cluster at the Weizmann Institute of Science. 
%\end{comment}


\end{document}